# Technological Parasitism

*Mario Coccia*

CNR -- National Research Council of Italy & Yale University School of Medicine

Current Address: Yale University School of Medicine
310 Cedar Street, Lauder Hall, Suite 118
New Haven, CT 06520

E-mail: mario.coccia@cnr.it

**Abstract**

Technological parasitism is a new theory to explain the evolution of technology in society. In this context, this study proposes a model to analyze the interaction between a host technology (system) and a parasitic technology (subsystem) to explain evolutionary pathways of technologies as complex systems. The coefficient of evolutionary growth of the model here indicates the typology of evolution of parasitic technology in relation to host technology: i.e., underdevelopment, growth and development. This approach is illustrated with realistic examples using empirical data of product and process technologies. Overall, then, the theory of technological parasitism can be useful for bringing a new perspective to explain and generalize the evolution of technology and predict which innovations are likely to evolve rapidly in society.

.

**Introduction**

This paper has two goals. The first is to propose a new perspective to measure and assess the evolution of technology, using a broad analogy with the evolutionary ecology of parasites. The second is to suggest properties that explain and generalize, whenever possible characteristics of the evolution of technology to predict which innovations are likely to evolve rapidly.

The analysis of the technology change and evolution of technology plays an important role in social studies of technology to explain the nature of innovation and predict patterns of technological innovation directed to solve problems and satisfy needs in society (Anadon et al., 2016; Andriani and Cohen, 2013; Angus and Newnham, 2013; Basalla, 1988; Freeman and Soete, 1987; Grodal et al., 2015; Hosler, 1994; Nelson and Winter, 1982; Rosenberg, 1969)[1]. In particular, measurement of the evolution of technology is an increasing challenge faced by governments, agencies and public research labs for improving technological forecasting and, as a consequence, supporting new technology for economic progress in society (cf., Coccia, 2005; Daim et al., 2018; Hall and Jaffe, 2018; Linstone, 2004; Tran and Daim, 2008). Scholars in this field of research endeavor of measuring technological advances of products and processes and technical performance of innovations with different approaches to explain determinants and directions of technological progress[2]. For instance, Nordhaus (1996, p. 29*ff*) applies an economic approach to estimate changes in lighting efficiency with a price index based on changes over the last two centuries, showing that the growth of real wages and real output in economic systems may have been significantly understated during the period since the Industrial Revolution. Other scholars apply engineering approaches to measure the advances of technical characteristics of innovations and explain different technological pathways (Dodson, 1985; Fisher and Pry, 1971; Knight, 1985; Martino, 1985; Sahal, 1981).

---

[1] cf. also, Abernathy and Utterback, 1978; Anderson and Tushman, 1990; Coccia, 2006, 2010, 2010a, 2011, 2014, 2014a, 2015, 2016, 2016a, 2017, 2017a, 2017b, 2018a, 2018b; Coccia and Wang, 2016; Henderson and Clark, 1990; Tushman and Anderson, 1986; Utterback, 1994.

[2] cf., Angus and Newnham, 2013; Coccia, 2005; Daim et al., 2018; Farrell, 1993; Farmer and Lafond, 2016; Faust, 1990; Koh and Magee, 2006, 2008; Magee et al., 2016; Nagy et al., 2013; Ruttan, 2001; Tran and Daim, 2008; Wang et al., 2016.

Although many studies of technology analysis, a technometrics that measures and assesses the evolution of technology as a complex system of interacting technologies is, at author's knowledge, unknown. The study here confronts this problem by developing a new approach to measure and assess the evolution of technology within theoretical framework of "Generalized Darwinism" (Hodgson and Knudsen, 2006, 2008). Wagner and Rosen (2014) argue that the application of evolutionary biology to different research fields has reduced the distance between life sciences and social sciences (cf., Nelson and Winter, 1982; Dosi, 1988). In general, analogies[3] derived from Darwinian evolutionary biology have provided meta-theoretical frameworks for interdisciplinary studies of the nature and evolution of technology (cf., Arthur, 2009; Arthur and Polak, 2006; Basalla, 1988; Coccia, 2018; Kauffman and Macready, 1995; Nelson, 2006). In fact, evolutionary biology, applied in economics of technical change, provides a logical structure of scientific inquiry to analyze and explain, in a broad analogy, characteristics and evolutionary pathways of technology (cf., Andriani and Cohen, 2013; Coccia, 2018; Wagner, 2011).

In general, technological change can be explained by a process of competitive substitution of a new technology for the old one (Fisher and Pry, 1971). However, technological progress is due to various aspects and dynamics of technological innovation (Coccia, 2005, 2018). Pistorius and Utterback (1997, p. 67) argue that a multi-mode interaction between technologies provides a much richer theoretical framework for technology analysis. In particular, Pistorius and Utterback (1997, p. 72ff) suggest different interactions among technologies in analogy with biology, more precisely: pure competition, symbiosis and predator-prey. Sandén and Hillman (2011, p. 407) discuss a further refinement of technological interactions by introducing a six-mode typology, using similarity with the interaction of species: neutralism, commensalism, amensalism, symbiosis, competition, and parasitism and predation into one category. A research challenge in this research field is the development of technometrics to measure different modes of technological interaction and transition between modes to explain the evolution of technology.

---

[3] cf., Oppenheimer, 1955.

In this context, the study here suggests a new conceptual framework for measuring and predicting technological evolution, applying a broad analogy with evolutionary ecology of parasites (cf., Coccia, 2018). In particular, the evolution of technology is analyzed here in simple way in terms of morphological changes between a host technology and a main technological parasitic subsystem. The proposed model assesses the types of interaction supporting the evolution of technology to suggest a technological forecasting of innovations that grow rapidly. This new perspective is verified on different technologies, using historical data. Overall, then, the theoretical framework here, borrowing conceptual insights from evolutionary ecology of parasites can extend the economics of technical change with a new approach that explains and generalizes evolutionary processes of innovation through interaction between technologies in a complex system. Moreover, results of this study here could aid policymakers and managers to predict which technologies are likely to evolve rapidly in order to design best practices of management of technology for accelerating industrial and economic change in society. In order to position this study within existing literature, next section describes different approaches for measuring technological advances.

**Theoretical background of the measurement of technological evolution**

The central issue for a theory of measurement is two basic problems: the first is the justification of assignment of the numbers to objects or phenomena (called *representational theorem*); the second is the specification of degree to which this assignment is unique (*uniqueness theorem*; cf., Luce et al. 1963; Suppes and Zinnes, 1963; Stevens, 1959). In the research field of the measurement of technology, technometrics is a theoretical framework for the measurement of technological advances and technological change with policy implications (Sahal, 1985; cf. also Sahal, 1981). Some approaches of the measurement of technological advances are described as follows, without pretending to present a comprehensive overview of the methods of technometrics (Coccia, 2005, p. 948*ff*).

*Hedonic approach*

The assumption of this approach is a positive relationship between market price of a good or service and its quality. In particular, it is assumed that a particular product can be represented by a set of characteristics and by their value; hence, the quality of product $Q_j$ is function of defining character-istics:

$$Q_j = f(a_1,...,a_n, X_{1j},...,X_{2j},...,X_{kj})$$

where $a_i$ is the relative importance of the $i$-th characteristics and $X_{ij}$ is the qualitative level of char-acteristics in product $j$. Technological progress can be defined here as the change in quality during a given period of time:

$$TC_j = \frac{\Delta Q_j}{\Delta t}$$

Moreover, the observed changes in the price of a product can be decomposed into a "quali-ty/technological change effect" and "pure price effect" (cf., Coccia, 2005, pp. 948-949; Saviotti, 1985, *passim*).

*RAND[4] approach*

A technological device has many technical parameters that measure its characteristics and charac-terize the state-of-the-art (SOA). Dodson (1985) proposes a planar and an ellipsoidal surface of SOA to measure technical advances of products:

*Planar*                                                     *Ellipsoidal*

$$\sum_{i=1}^{n} \left( \frac{x_i}{a_i} \right) = 1 \qquad\qquad\qquad \sum_{i=1}^{n} \left( \frac{x_i}{a_i} \right)^2 = 1$$

where $x_i$ is the $i$-th technological characteristic and $a_i$ is the $i$-th parameter (a constant). Alexander and Nelson (1973) suggest hyperplanes for the surface of SOA, instead of ellipses. In brief, Hedonic and RAND techniques for measuring technological advances are similar and differ only in the choice of dependent variable, which is price in the former and calendar year in the latter (Coccia, 2005, pp. 949-952).

*Functional and Structural approach*

The technique by Knight (1985) is based on a functional and a structural description of a given technology to detect its evolution. In regard to the functional description of a new computer over an earlier one, this technique can indicate how technological advancement has taken place, but it does not specify the details of new development. In order to explain technological issues, it is also necessary the structural description between technologies by comparing the structure of new systems with that of earlier ones (cf., Coccia, 2005, pp. 955-957). The structural approach was originated by Burks et al. (1946) that describe the "logical design for a general-purpose digital computer", showing key information needed to determine its functional performance and computing power (as quoted by Knight, 1985, p. 109).

*Wholistic and Holistic approaches*

Sahal (1985) suggests two ideas of technometrics. In the first approach called *Wholistic*, the state-of-the-art (SOA) is specified in terms of a surface of constant probability density given the distribution of technological characteristics. The SOA at any given point in time is represented by a probability mountain, rising above the geometric plane. The level of technological capability is given by the height of mountain. Instead, the magnitude of technological change can be estimated by the difference in heights of successive mountains. In the second approach called *Holistic*, a technological characteristic is specified as a vector in an *N*-dimensional space generated by a set of *N* linearly independent elements, such as mass, length, and time. The length of the vector represents the magnitude of a technological characteristic, whereas the type of characteristic is represented by direction. In this case, the SOA reduces to a point. The successive points at various times constitute a general pattern of technological evolution that evinces a series of *S*-shaped curves. These two approaches are distinct but related (Coccia, 2005, p. 955).

*Model of technological substitution for measuring technological evolution*

Fisher and Pry (1971, p. 75) argue that technological evolution consists of substituting a new technology for the old one, such as the substitution of coal for wood, hydrocarbons for coal, robotics technologies for humans (cf., Daim et al., 2018), etc. Technological advances are represented by competitive substitutions of one method of satisfying a need for another. Fisher and Pry (1971, p. 88) state that: "The speed with which a substitution takes place is not a simple measure of the pace of technical advance . . . . it is, rather a measure of the unbalance in these factors between the competitive elements of the substitution".

---

[4] RAND Corporation ("Research ANd Development") is an U.S. research organization that develops researches to support the security, health and economic growth of the USA, allied countries and in general the world.

*Technological advances measured with patent data*

Faust (1990, p. 473) argues that patent indicators allow for a differentiated observation of technological advances before the actual emergence of an innovation, such as technological development in the scientific field of superconductivity. Wang et al. (2016, p. 537ff) investigate technological evolution using US Patent Classification (USPC) reclassification. Results suggest that: "patents with Inter-field Mobilized Codes, related to the topics of 'Data processing: measuring, calibrating, or testing' and 'Optical communications', involved broader technology topics but had a low speed of innovation. Patents with Intra-field Mobilized Codes, mostly in the Computers & Communications and Drugs & Medical fields, tended to have little novelty and a small innovative scope" (Wang et al., 2016, p. 537, original emphasis). Future research in this research field should extend the patent sample to subclasses or reclassified secondary USPCs in order to explain in-depth technological evolution within a specific scientific field.

*Other approaches for measuring technological evolution*

New criteria of technological assessment apply technology development envelope (extension of hierarchical decision modeling and analytical hierarchy process into the future) to detect multiple pathways for technological evolution and construct strategic roadmapping, as illustrated by Daim et al. (2018, p. 49ff) for robotics technologies.

Koh and Magee (2006, 2008) suggest an approach for studying technological progress based on three functional operations—storage, transportation and transformation. Results for information and energy technology indicate a continuous progress for each functional category independent of the specific underlying technological artifacts dominating at different times. However, some differences between energy and information technology are seen (cf. also, Valverde, 2016 for transitions in in-formation technology). Magee et al. (2016) show that Moore's law is a better description of long-term technological change when the performance data come from various designs, whereas experience curves may be more relevant when a singular design in a given factory is considered. In particular, Magee et al. (2016, p. 245) argue that: "Moore's exponential law appears to be more fundamental than Wright's power law for these 28 domains (where performance data are record breakers from numerous designs and different factories)". Moreover, Wright's approach shows that the cost of technology decreases as a power law of cumulative production, whereas generalized Moore's law shows that technologies improve performance exponentially with time. Nagy et al. (2013, p. 1)–using a statistical model to rank the performance of the postulated laws applied on cost and production of 62 different technologies–claim that:

> Wright's law produces the best forecasts, but Moore's law is not far behind …. results show that technological progress is forecastable, with the square root of the logarithmic error growing linearly with the forecasting horizon at a typical rate of 2.5% per year. These results have implications for theories of technological change, and assessments of candidate technologies and policies for climate change mitigation.

In this context, for predicting technological progress, Farmer and Lafond (2016, p. 647): "formulate Moore's law as a correlated geometric random walk with drift, and apply it to historical data on 53 technologies….. to make forecasts for any given technology with a clear understanding of the quality of the forecasts. … to estimate the probability that a given technology will outperform another technology at a given point in the future".

Table 1. Strengths and weaknesses of some technometric approaches

| Technometrics | Strengths | Weaknesses |
|---|---|---|
| Hedonic | Hedonic function estimates a price surface.<br><br>Hedonic method considers both economic and technical information. | First, the technique works best in cases of a distinct product technology. It cannot easily be applied to cases of a process technology.<br><br>Second, the Hedonic approach is unsuitable for international comparisons because of significant differences in factor prices among countries.<br><br>Third, it cannot be used in an 'unskilled' way to measure changes in technology .<br><br>Finally, its theoretical status is still not clear. |
| RAND | State of the art (SOA) surfaces can reveal whether technological changes are "biased" toward increasing the relative availability (decreasing the relative cost) of one characteristic, or a group of them, relative to others. | First, the estimation procedure is arbitrary and difficult.<br><br>Second, it does not take into account the correlations between technological characteristics, thereby seriously obscuring if not distorting the real rate and extent of technical progress. |
| Functional and Structural | The methodology has excellent potential application for most product and production technologies. | The full use of the functional/structural analysis to isolate and describe specific technologic advances and their values has found limited successful use. |
| Wholistic and Holistic | Wholistic. The framework provides an objective basis for determining the critical variables in the evolution of technology. The reproducibility of the results is excellent.<br><br>Holistic. It provides an *a priori* theoretical basis for the selection of relevant variables, the choice of a functional form, and the quantification of weights assigned to each of the variables. It is possible to identify the sources underlying the observed advances in technology. | Methodological issues (e.g., data collection, etc.). |
| Fisher and Pry's Model | Technological advances are represented by competitive substitutions between new and old products. | Technological progress is due to multi-mode interaction among technologies rather than mere competition. |

Table 1 synthetizes some approaches of the measurement of technological advances with pros and cons. Many techniques of the analysis of technological advances focus on competition between technologies, such as substitution model by Fisher and Pry (1971) and predator-prey interaction by Pistorius and Utterback (1997). This study here endeavors to measure the evolution of technology considering an alternative perspective based on interactions between a host-master technology and its main parasitic subsystem of technology to predict long-term development of the complex system of technology (cf., Coccia, 2018). Next section presents the conceptual framework of the suggested technometrics here.

**Evolutionary ecology of technology within a Generalized Darwinism**

The scientific departure of the proposed technometrics here is principles of the "Generalized Darwinism" (Hodgson and Knudsen, 2006, 2008) that provide suitable concepts for framing a broad analogy between evolution of technologies and evolutionary ecology of parasites to measure and explain different evolutionary pathways of technology itself. In economics of technical change, the generalization of Darwinian principles ("Generalized Darwinism") can assist in explaining the multidisciplinary nature of many innovation processes (cf., Basalla, 1988; Farrell, 1993; Hodgson and Knudsen, 2006; Levit et al., 2011; Nelson, 2006; Schubert, 2014; Wagner and Rosen, 2014). In this context, Arthur (2009) argues that sociocultural evolution is related to the evolution of technology and Darwinism can explain technology development as it has done for species development (cf., Schuster, 2016, p. 7). In general, technological evolution, as biological evolution, displays radiations, stasis, extinctions, and novelty (Valverde et al., 2007). Kauffman and Macready (1995, p. 26, original emphasis) state that: "Technological evolution, like biological evolution, can be considered a search across a space of possibilities on complex, multipeaked 'fitness,' 'efficiency,' or 'cost' landscapes". Schuster (2016, p. 8) argues that: "Technologies form complex networks of mutual dependences just as the different species do in the food webs of ecosystems". Kauffman and Macready (1995, p. 27 and p. 42) also point out that:

> Evolution, biological or technological, is actually a story of coevolution. Adaptive alterations by the predatory bat alter the adaptive landscape of its frog prey. Alterations in the maximum power of the engine of an automobile alter optimal tire, suspension, and even highway design. Coevolution is a process of coupled, deforming landscapes where the adaptive moves of each entity alter the landscapes of its neighbours in the ecology or technological economy (p.27)....Biological and technological evolution are both characterized by the requirement to solve hard combinatorial optimization problems. . . . These interrelated features of many hard combinatorial optimization problems are therefore likely to underlie features of biological and technological evolution (p.42).

Nelson (2006, p. 491) claims that a broad approach of Universal Darwinism in social sciences is: "a roomy intellectual tent welcoming scholars studying a variety of topics".

The crux of the study here is to measure and assess the evolution of technologies in a broad analogy with evolutionary ecology of parasites within a setting of Generalized Darwinism. Some brief backgrounds of the evolutionary ecology of parasites are useful to clarify the technometrics proposed here. Firstly, ecology studies the relationship functions and interactions between organisms of the same or different species and environment in which they live (cf., Poulin, 2006). In particular, the scope of the ecology is to explain all sorts of interaction of organisms to one another and to their environment. Secondly, the evolutionary ecology of parasites focuses on parasites (from Greek *para* = near; *sitos* = food) that are any life form finding their ecological niche in another living system (host). Parasites have a range of traits that evolve to locate in available hosts, survive and disperse among hosts, reproduce and persist (cf., Janouskovec and Keeling, 2016). Coccia (2018) argues that technologies can have a behavior similar to parasites because technologies cannot survive and develop as independent systems *per se*, but they can function and evolve in markets if associated with other host technologies, such as audio headphones, speakers, software apps, etc. that function if and only if they are associated with host electronic devices (e.g., smartphone, radio receiver, television, etc.).

This study endeavors to measure the effect that one host technology has on growth rate of parasitic technology to explain the evolution of the overall complex system of technology.

**Model for the evolution of technology in complex systems**

Evolution is a stepwise and comprehensive development [it originates from Latin *evolution –onis,* der. of *evolvĕre* = act of carrying out (the papyrus)]. In general, the process of development generates the formation of complex systems in nature and society (cf., Barton, 2014). The theoretical framework of "Universal Darwinism" (Dawkins, 1983; Nelson, 2006) claims that: "Darwinism involves a general theory of all open, complex systems" (Hodgson 2002, p. 260; cf., Levit et al., 2011). Hodgson and Knudsen (2006) suggest a generalization of the Darwinian concepts of selection, variation and retention to explain how complex systems evolve (cf. also, Hodgson, 2002; Stoelhorst, 2008). Hence, in order to show the proposed metrics of the evolution of technology here, it is important to clarify the concept of complex system. Simon (1962, p. 468) in the study of complexity states that: "a complex system [is]… one made up of a large number of parts that interact in a nonsimple way …. complexity frequently takes the form of hierarchy, and …. a hierarchic system … is composed of interrelated subsystems, each of the latter being, in turn, hierarchic in structure until we reach some lowest level of elementary subsystem." In the field of technology, McNerney et al. (2011, p. 9008) argue that: "The technology can be decomposed into $n$ components, each of which interacts with a cluster of $d - 1$ other components" (cf., Gherardi and Rotondo, 2016; Oswalt, 1976; Magee, 2012, p. 16ff. for materials innovation). Arthur (2009, pp. 18-19) claims that: "Technologies somehow must come into being as fresh combinations of what already exists". This combination of components and assemblies is organized into systems to some human purpose and has a hierarchical and recursive structure. In particular, the evolution of technology is due to major innovations and numerous minor innovations that interact in a complex system of technology (cf., Coccia, 2018; Sahal, 1981, p. 37). Sahal (1981) points out that: "evolution…pertains to the very structure and function of the object (p. 64) …. involves a process of equilibrium governed by the internal dynamics of the object system (p. 69)". Moreover, the short-term evolution of technology is due to changes within system, whereas the long-term evolution is possible by forming an integrated system (Sahal, 1981, pp. 73-74). This study here endeavors, starting from theoretical background discussed above, to measure and assess interaction between technologies within a host-parasite system for forecasting evolutionary pathways over time[5]. The following premises support the technometrics here (Coccia, 2018):

◻ Technology is a complex system composed of more than one entity or sub-system and a relationship that holds between each entity and at least one other entity in the system. The technology is adapted in the Environment $E$ with a natural selection operated by market forces and/or artificial selection operated by human beings (based on efficiency, technical,

---

environmental and economic characteristics) to satisfy needs, achieve goals and/or solve problems in human society.

- ☐ In the long run, the behavior and evolution of any technology is not independent from the behavior and evolution of the other technologies (Coccia, 2018).

- ☐ Interaction between technologies is an interrelationship of information/resources/energy and other physical/chemical phenomena for reciprocal adaptations in inter-related complex systems.

- ☐ Coevolution of technologies is the evolution of reciprocal adaptations in a complex system supporting the reciprocal enhancement of technologies' growth rate—i.e., a modification and/or improvement of technologies based on interaction and adaptation in complex systems and markets to satisfy changing needs and solve consequential problems in society.

- ☐ *P* is a parasitic technology in *H* (host or master technology) if and only if during its life cycle, technology P is able to interact and adapt into the complex system of technology H, generating coevolutionary processes to satisfy needs, achieve goals, and/or solve problems in society.

In general, technologies form complex systems based on subsystems of technology that interact in a non-simple way (e.g., batteries and antennas in electronic devices; cf., Coccia, 2018). Overall, then, the interaction between technologies in a complex system tends to generate stepwise coevolutionary processes within "space of the possible" (Wagner and Rosen, 2014, *passim*).

In order to operationalize the approach here to measure, assess and predict the evolution of technology here, this study proposes a simple model of technological interaction between a host technology *H* and an interrelated parasitic subsystem of technology. This model measures changes in a subsystem of parasitic technology in relation to proportional changes in the overall host system of technology. In particular, this model measures the effect that one host technology has on parasitic technology's growth rate. This approach is based on the biological principle of allometry that was originated to study the differential growth rates of the parts of a living organism's body in relation to the whole body (cf., Reeve and Huxley, 1945 for evolutionary biology studies; Sahal, 1981 for patterns of technological innovation).

The general model is based on following assumptions.

(1) Suppose the simplest possible case of only two technologies, $H$ (a host or master technology) and $P$ (a parasitic subsystem of technology in $H$), forming a Complex System $S$(H, P); of course, the model can be generalized for complex systems including many subsystems of technology.

(2) Let $P(t)$ be the extent of technological advances of a technology $P$ at the time $t$ and $H(t)$ be the extent of technological advances of a technology $H$ (master or host system) that interacts with $P$ at the same time (cf., Sahal, 1981, pp. 79-89). Suppose that both $P$ and $H$ evolve according to some $S$-shaped pattern of technological growth, such a pattern can be represented analytically in terms of the differential equation of logistic function. For $H$, Host technology, the starting equation is:

$$\frac{1}{H}\frac{dH}{dt} = \frac{b_1}{K_1}\left(K_1 - H\right)$$

The equation can be rewritten as:

$$\frac{K_1}{H}\frac{1}{\left(K_1 - H\right)}dH = b_1 dt$$

The integral of this equation is:

$$\log H - \log\left(K_1 - H\right) = A + b_1 t$$

$$\log\frac{K_1 - H}{H} = a_1 - b_1 t$$

$$H = \frac{K_1}{1 + \exp\left(a_1 - b_1 t\right)}$$

$a_1 = b_1 t$ and $t$ = abscissa of the point of inflection.

The growth of $H(t)$ can be described respectively as:

$$\log\frac{K_1 - H}{H} = a_1 - b_1 t \qquad\qquad [1]$$

*Mutatis mutandis*, for Parasitic technology $P(t)$ the equation is:

$$\log \frac{K_2 - P}{P} = a_2 - b_2 t \qquad [2]$$

The logistic curve here is a symmetrical *S*-shaped curve with a point of inflection at 0.5K with $a_{1,2}$ are constants depending on the initial conditions, $K_{1,2}$ are equilibrium levels of growth, and $b_{1,2}$ are rate-of-growth parameters (1=Host technological system, 2=Parasitic technological subsystem).

Solving equations [1] and [2] for *t*, the result is:

$$t = \frac{a_1}{b_1} - \frac{1}{b_1} \log \frac{K_1 - H}{H} = \frac{a_2}{b_2} - \frac{1}{b_2} \log \frac{K_2 - P}{P}$$

The expression generated is:

$$\frac{H}{K_1 - H} = C_1 \left( \frac{P}{K_2 - P} \right)^{\frac{b_1}{b_2}} \qquad [3]$$

Equation [3] in a simplified form is $C_1 = exp[b_1(t_2 - t_1)]$ with $a_1 = b_1 t_1$ and $a_2 = b_2 t_2$ (cf. Eqs. [1] and [2]); when *P* and *H* are small in comparison with their final value, the model of technological evolution of the host-parasite system is given by:

$$P = A \ (H)^B \qquad [4]$$

where $A = \frac{K_2}{(K_1)^{\frac{b_2}{b_1}}} C_1$ and $B = \frac{b_2}{b_1}$

The logarithmic form of the equation [4] is a simple linear relationship:

$$\log P = \log A + B \ \log H \qquad [5]$$

$B$ is the evolutionary coefficient of growth that measures the evolution of technology $P$ (Parasite) in relation to $H$ (Host or Master technology).

This model of the evolution of technology [5] has linear parameters that are estimated with the Ordinary Least-Squares Method. The value of $B \gtrless 1$ in the model [5] measures the relative growth of $P$ in relation to the growth of $H$ and it indicates different patterns of technological evolution: B<1

(underdevelopment), B ≥ 1 (growth or development of technology P). In particular,

☐ $B < 1$, whether technology $P$ (a subsystem of $H$) evolves at a lower relative rate of change than technology $H$; the whole system of technology $S$(H, T) has a slowed evolution (*underdevelopment*) over the course of time.

☐ $B$ has a unit value: $B = 1$, then the two technologies $P$ and $H$ have proportional change during their evolution: i.e., a symmetrical coevolution between a system of technology ($H$) and its interacting subsystem P. In short, when B=1, the whole system of technology $S$(H, T) here has a proportional evolution (*growth*) of its sub-systems of technology.

☐ $B > 1$, whether $P$ evolves at greater relative rate of change than $H$; this pattern denotes disproportionate technological advances in the structure of a subsystem $P$ as a consequence of change in the overall structure of a host technological system $H$. The whole system of technology S(H,T) has an accelerated evolution (*development*) over the course of time.

The coefficient $B$ of evolutionary growth can be a metric for classifying the modes of interaction between technologies. Moreover, this coefficient $B$ is systematized in an ordinal scale that indicates typologies of the evolution of technology and grade of how a host technology can enhance or inhibit the growth rate of parasitic technology (table 2).

Table 2. Scale of the evolution of technological subsystem *P* in relation to Host technology *H*

| Grade of evolution of the system of technology | Coefficient of evolutionary growth of the subsystem of technology P | Type of the evolution of subsystem of technology P in relation to H (Symbol) | Mode of technological interactions between technologies H and P | Evolution of overall complex system of technology (Symbol) | Predictions of the evolution of overall system of technology |
|---|---|---|---|---|---|
| 1 Low | B<1 | Reduced / | **Parasitism** | Underdevelopment / | Complex system of technology evolves slowly over time |
| 2 Average | B=1 | Proportional + | **Mutualism** | Growth + | Complex system of technology has a steady-state growth |
| 3 High | B>1 | Accelerated ! | **Symbiosis** | Development ! | Complex system of technology is likely to evolve rapidly |

*Note*: Symbols /, +, ! indicate in brief the type of technological evolution: underdevelopment, growth and development respectively.

Table 2 also suggests some symbols to indicate the intensity of growth rate of complex system of technology, measured with the coefficient of evolutionary growth *B* in model [5]: \ = underdevelopment, +=growth, and != development.

Properties of the scale of the evolution of technology are (table 2):

a) Technology of higher rank-order on the scale (with B>1) has higher technological advances of lower rank-order technologies (with B<1).

b) If a technology has the highest ranking on the scale (i.e., with B>1), it evolves rapidly (development) over the course of time. *Vice versa*, if a technology has the lowest ranking on the scale (with *B*<1), it evolves slowly (underdevelopment).

c) Technology of the highest rank order on the scale (with *B*>1) has accumulated all previous evolutionary stages of low rank order and generates a symbiotic growth between a system of technology *H* and its interacting subsystem of technology *P*.

d) The logical relation of interactions between technologies is: technological parasitism ⊆ technological mutualism ⊆ technological symbiosis (the symbol ⊆ indicates subset in the set theory).

The model here suggests different grades of technological evolution of the subsystem of technology *P* supporting the evolution of overall complex system of technology. In particular, the initial stage of technological interaction is a technological parasitism between host and parasitic subsystem of technology (*B*<1). The change of coefficient *B* indicates the shift towards modes of stronger interaction between technologies within a complex system, such as technological mutualism (*B*=1) and technological symbiosis (*B*>1) that lead to a coevolution of the overall system of technology (cf., Coccia, 2018). Hence,

☐ *B*<1 indicates mainly a *Technological parasitism:* any type of relationships between two technologies where one technology *P* (subsystem technology) benefits from the other (Host) that, instead, has a negative benefit from this interaction. This relationship can generate a low development of the subsystem technology and, as a consequence, of the overall complex system of technology (cf., Coccia, 2018). The low growth of the complex system of technology is due to an unidirectional and asymmetrical effect from H →P

☐ *B*=1 indicates a *Technological mutualism:* any type of relationships in which each technology benefits from the activity of the other technology. This interaction between technologies supports mutual benefits with symmetric and proportion evolutionary growth both of host system of technology *H* and of parasitic subsystem of technology *P*. The bi-directional relation of growth is given by: *H* ↔ *P*.

☐ *B*>1 indicates a *Technological symbiosis:* any type of long-term relationships between technologies that interact and evolve together in a complex system. The technological interaction between *H* and *P* is: *H* ⇔(*strong*) *P*.

**Materials and method**

*Data and their sources*

The evolution of technology is measured here using historical data of five example technologies (four for US market and one for Italian market); farm tractor, freight locomotive, generation of electricity in steam-powered and internal-combustion plants in the United States of America. In fact, US national system of innovation is a vital case study that shows general patterns of the evolution of technology across advanced market economies (Steil et al., 2002). Sources of data for these technologies are tables published by Sahal (1981, pp.319-350, originally sourced from trade literature; cf. also Coccia, 2018). Note that data from the earliest years and also the war years are sparse for some technologies. In addition, this study also considers data of a main Information and Communication Technology (ICT): smartphone. Data of smartphone here are originally sourced from trade literature of Italian market, one of the largest economy in Europe (Punto Cellulare, 2018). Historical data of these technologies are important to verify applicability, effectiveness, generality, precision, correctness and robustness of the proposed model of technological evolution.

*Measures*

Functional Measures of Technology (FMT) are the technical characteristics of innovations and their change can indicate the evolution of technology over the course of time based on major and minor innovations, such as fuel-consumption efficiency of vehicles (cf., Sahal, 1981, pp. 27-29). The following FMTs are associated with a main subsystem of technology that indicates a parasitic technology $P$, and a host system $H$ in which the parasitic technology $P$ operates and interacts. FMTs per each technology seem to be the most appropriate variables to apply the suggested model of host-parasitic system for measuring and predicting the evolution of technology. Other measures are not considered here because they do not provide complete information of technical characteristics of technologies under study, such as index of tractor price in relation to price of labor, number of locomotive in service, cumulated production quantities, etc.

1. Functional Measures of Technologies (FMTs) for farm tractor over 1920-1968 CE (Common Era) in US market are:

– fuel-consumption efficiency in horsepower-hours indicates the technological advances of engine (a parasitic technology $P$) within farm tractors. This FMT represents the dependent variable $P$ in the model [5].

– mechanical efficiency (ratio of drawbar horsepower to belt or power take-off –PTO- horsepower) is a proxy of the technological advances of farm tractor ($H$=Host technology). This FMT represents the explanatory variable $H$ in the model [5].

2. For freight locomotive, FMTs over 1904-1932 CE in US market are:

– Tractive efforts in pound indicate the technological advances of locomotive (Parasitic technology P). This FMT represents the dependent variable $P$ in the model [5].

– Total railroad mileage indicates the evolution of the infrastructure system of railroad (Host technology). This FMT represents the explanatory variable $H$ in the model [5].

3. For electricity generated by steam-powered plants, FMTs over 1920-1970 CE in US market are:

– Average fuel-consumption efficiency in kilowatt-hours per pound of coal indicates the technological advances of boiler, turbines and electrical generator (parasitic technology P of steam-powered plant). This FMT represents the dependent variable $P$ in the model [5].

– Average scale of plant utilization (the ratio of net production of steam-powered electrical energy in millions of kilowatt-hours to number of steam powered plants) indicates a proxy of technological advances of the steam-powered plant (Host technology). This FMT represents the explanatory variable $H$ in the model [5].

4. For electricity generated by internal-combustion plants, FMTs over 1920-1970 CE in US market are:

– Average fuel-consumption efficiency in kilowatt-hours per cubic foot of gas indicates the technological advances of boiler, turbines and electrical generator (a parasitic subsystem of internal combustion plant). This FMT represents the dependent variable $P$ in the model [5].

– Average scale of plant utilization (the ratio of net production of electrical energy by internal-combustion type plants in millions of kilowatt-hours to total number of these plants) indicates a proxy of technological advances of plants with internal-combustion technology. This FMT represents the explanatory variable of the host technology *H* in the model [5].

5. This study also considers smartphone technologies by using a sample of *N*=738 models of famous brands (Apple, ASUS, HTC, Huawei, LG Electronics, Motorola, Nokia, Samsung, Sony, ZTE, etc.) from 2008 to 2018, sold in Italy during the years 2012 and 2018. Functional Measures of Technological characteristics (FMTs) in smartphone technology over 2008-2018 CE in Italian market are given by:

− Main Camera in megapixel (Mpx) indicates the technological advances of camera technology (Parasitic technology *P*) in smartphone. This FMT represents the dependent variable *P* in the model [5].

− Processor GHz (Giga Hertz, GHz) indicates a proxy of the technological advances of overall smartphone technology (Host technology *H*). This FMT represents the explanatory variable *H* in the model [5].

In addition, in order to assess the multidimensional process of interaction between host technology and parasitic technologies, this case study of smartphone technology also considers further FMTs over 2008-2018 period given by:

− Display resolution in total pixels[6]= display size row × display size column

− Second Camera Mpx (megapixel)

− Memory Gb (Giga byte)

− RAM Gb (Giga byte)

− Battery mAh (milliAmpere hour)

*Model and data analysis procedure*
Model [5] of the technological evolution is specified as follows:

*log $P_t$ = log a + B log $H_t$ + $u_t$*                    [6]

*a* is a constant; *log* has base *e*= 2.7182818; *t*=time; *$u_t$* = error term .

$P_t$ will be the extent of technological advances of technology $P$ (a parasitic subsystem of the Host technology $H$ at time $t$).

$H_t$ will be the extent of technological advances of host technology $H$ in which the parasitic subsystem of technology $P$ interacts at time $t$; $H$ technology as a complex system is the driving force of the evolutionary growth of overall interrelated subsystems of technology $P_i$ ($i$=1, . . . , n).

The multidimensionality is considered with the following model:

$$log\ P_{1t} = log\ a + B_1\ log\ H_t + B_2\ log\ P_{2t} + B_i\ log\ P_{it} + ... + B_m\ log\ P_{mt} + \varepsilon_t \qquad [7]$$

$H_t$=Host technology; $P_{it}$= Parasitic technology $i$ ($i$=1, ..., n); $t$=time; $\varepsilon_t$ = error term.

The equations of simple regression [6] and multiple regression [7] are estimated using the Ordinary Least Squares method. Statistical analyses are performed with the Statistics Software SPSS® version 24.

**Case studies of the evolution of technology in agriculture, rail transport, electricity generation and smartphone**

*Results of the evolution of farm tractor technology (1920-1968 period in US market)*

Table 3 shows that the evolutionary coefficient of growth of farm tractor technology, from model [6], is $B$ = 1.74, i.e., $B$ >1: the subsystem technology of engine ($P$) has a disproportionate technological growth in comparison with overall farm tractor ($H$). This coefficient indicates a high grade of the evolution of technology $P$ and a development of the whole system of farm tractor technology (cf., Figure 1).

---

Table 3 – Estimated relationship for farm tractor technology (1920-1968 period in US market)

*Dependent variable*: *log* fuel consumption efficiency in horsepower hours (*P*=technological advances of engine within tractor)

|  | Constant $\alpha$ (St. Err.) | Evolutionary coefficient $\beta=B$ (St. Err.) | $R^2$ adj. (St. Err. of the Estimate) | F (sign.) |
|---|---|---|---|---|
| Farm tractor | −5.14*** | 1.74*** | 0.85 | 256.44 |
|  | (0.45) | (0.11) | (0.10) | (0.001) |

*Note*: ***Coefficient β is significant at 1‰; Explanatory variable is *log* mechanical efficiency ratio of drawbar horsepower to belt (technological advances of farm tractor –Host technology *H*)

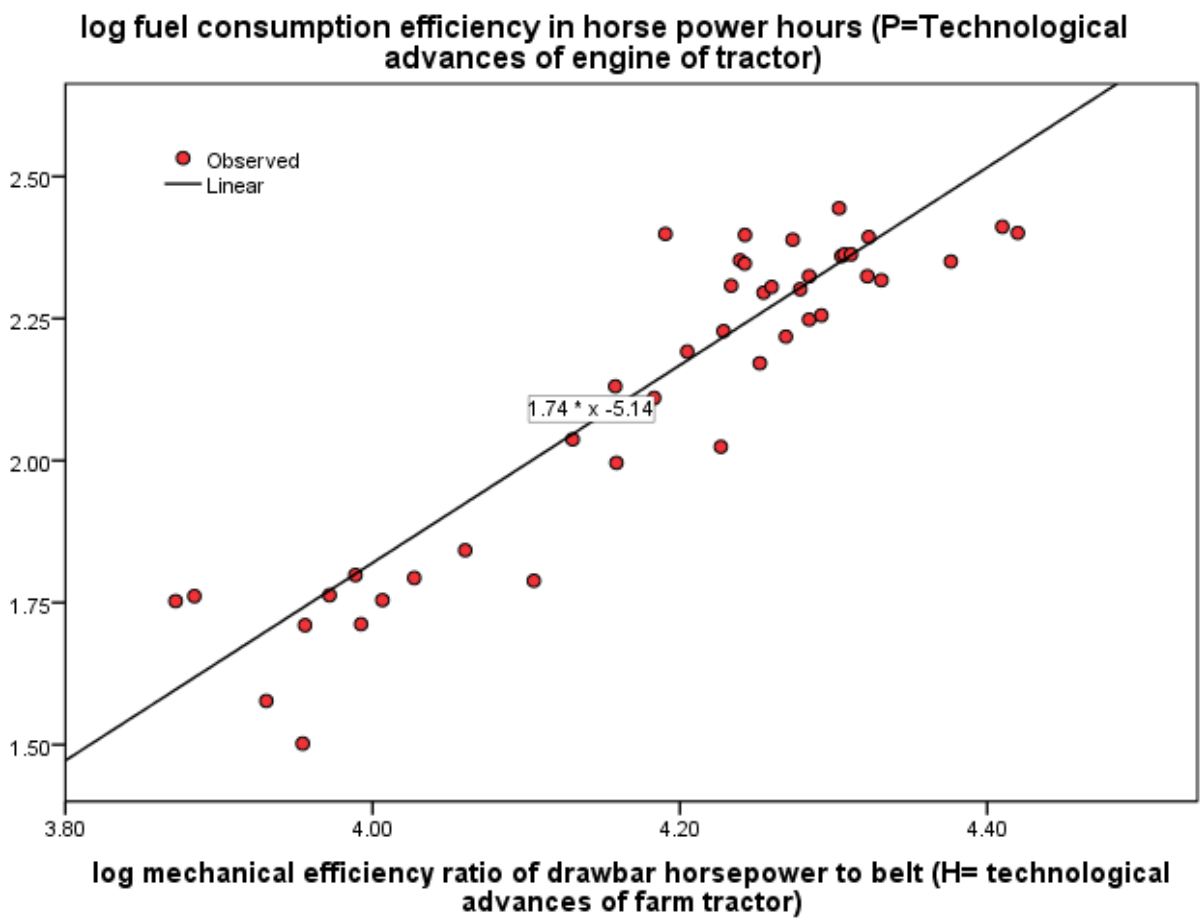

**Figure 1.** Trend and estimated relationship of the evolution of farm tractor technology (1920-1968 period in US market)

This result confirms the study by Sahal (1981) that the rapid evolution of farm tractor technology is due to numerous incremental and radical innovations over time, such as the diesel-powered track-type tractor in 1931, low-pressure rubber tires in 1934 and the introduction of remote control in 1947 that made possible improved control of large drawn implements. The development of the continuous running power takeoff (PTO) also in 1947 allowed the tractor's clutch to be disengaged without impeding power to the implements. Moreover, in 1950 it is introduced the 1000-rpm PTO for transmission of higher power, whereas in 1953 power steering was applied in new generations of tractor. In addition, the PTO horsepower of tractor has more than doubled from about 27hp to 69hp over 1948-1968; finally, dual rear wheels in 1965, auxiliary front-wheel drive and four-wheel drive in 1967 have improved the overall technological performance of tractor (Sahal, 1981, p. 132ff). These radical and incremental innovations have supported the accelerated evolution of farm tractor technology over time as confirmed by the statistical evidence here with the coefficient of evolutionary growth B>1 (grade 3=high in table 2).

*Results of the evolution of freight locomotive technology (1904–1932 period in US market)*

Table 4 shows that the evolutionary coefficient of freight locomotive technology is $B = 1.89$, i.e., $B> 1$: this coefficient of growth indicates a process of development of freight locomotive technology P in the host system of rail transportation (see, Figure 2).

Table 4 – Estimated relationship for freight locomotive technology
(1904–1932 period in US market)

| *Dependent variable*: *log* Tractive efforts in pound (*P*=technological advances of locomotive) | | | | |
|---|---|---|---|---|
| | Constant $\alpha$ (St. Err.) | Evolutionary coefficient $\beta=B$ (St. Err.) | $R^2$ adj. (St. Err. of the Estimate) | F (sign.) |
| Locomotive technology | −13.87*** | 1.89*** | 0.91 | 270.15 |
| | (1.48) | (0.12) | (0.07) | (0.001) |

*Note*: ***Coefficient β is significant at 1‰; Explanatory variable is *log* Total railroad mileage (technological advances of the infrastructure –Host technology *H*)

This development of freight locomotive technology can be explained with a number of technological advances, such as the introduction of compound engine in 1906 that improved tractive effort (Sahal, 1981). In 1912 the first mechanical stoker to use steam-jet overfeed system of coal distribution was perfected. In 1913, another technological advance was the substitution of pneumatically operated power reverse gear for the hand lever. In 1916, the introduction of the unit drawbar and radial buffer eliminated the need for a safety chain in coupling the engine and tender together. Fur-ther technological advances are due to the adoption of cast-steel frames integral with the cylinder, the chemical treatment of the locomotive boiler water supply and the introduction of roller bearings over 1930s. In particular, these technical developments reduced the frequency of maintenance work in locomotives. Subsequently, the continuous modification of steam locomotive with reciprocating engine has led to diesel-electric locomotive by the mid-1940s (Sahal, 1981, p. 154*ff*). These and other technological developments have supported the accelerated evolution of freight locomotive technology over time as confirmed by the coefficient of evolutionary growth B>1 calculated in table 4.

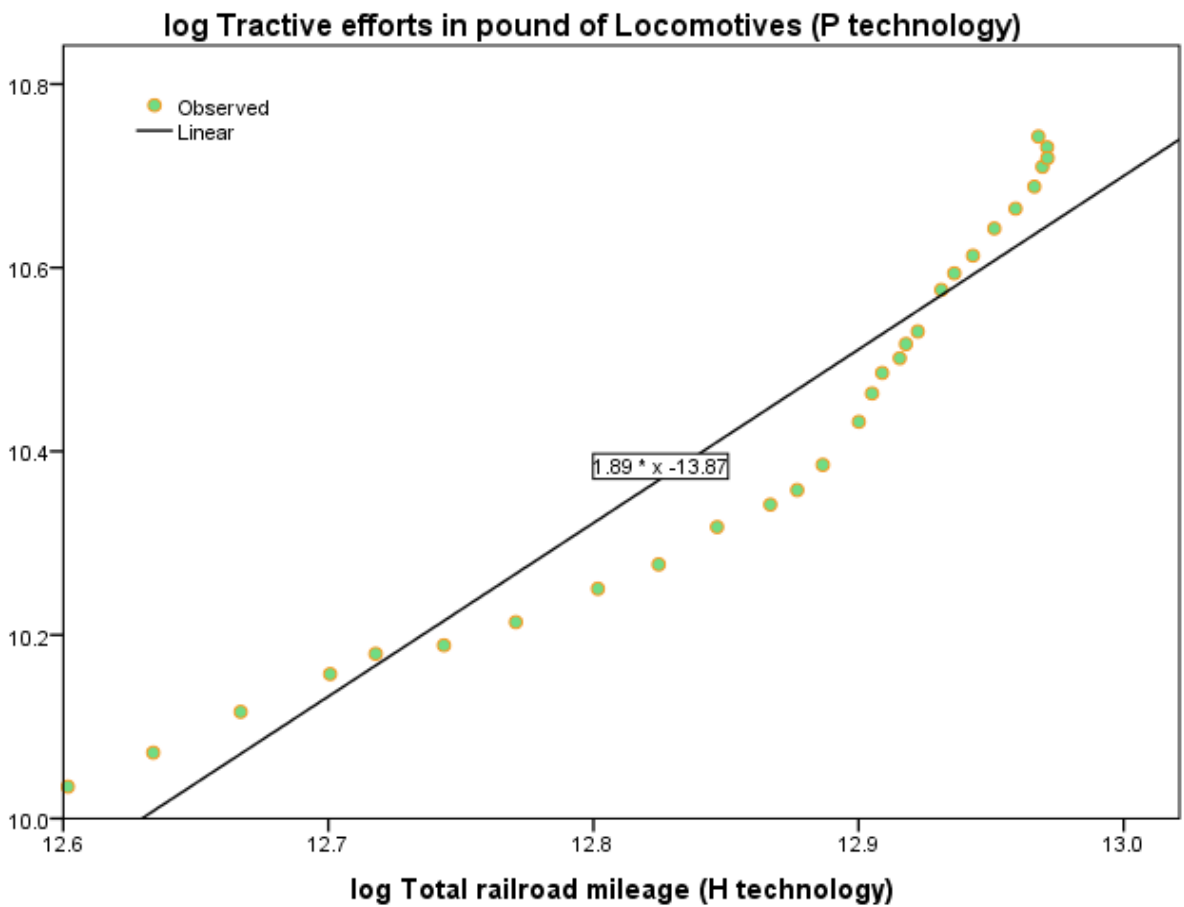

**Figure 2.** Trend and estimated relationship of the evolution of freight locomotive technology (1904–1932 period in US market)

*Results of the evolution of electricity generation technology (1920-1970 period in US market)*

Electricity is generated in different types of plants: 1. Steam-powered plants, which may be either fossil fueled or nuclear plant; 2. Internal-combustion plants, including gas turbines and diesel en-gines; 3. hydroelectric plants. This study focuses on 1[st] and 2[nd] type of plants. Table 5 shows that the steam-powered electricity, with plants that are fossil (coal) fueled, has $B = 0.23$, i.e., $B < 1$ (see also Figure 3).

Table 5 – Estimated relationship for steam-powered plants that are fossil (coal) fueled
(1920-1970 period in US market)

| *Dependent variable*: *log* Average fuel consumption efficiency in kwh per pound of coal (*P*=technological advances of turbine and various equipment) | | | | |
|---|---|---|---|---|
| | *Constant* $\alpha$ *(St. Err.)* | *Evolutionary Coefficient $\beta=B$ (St. Err.)* | $R^2$ *adj. (St. Err. of the Estimate)* | *F (sign.)* |
| Turbine and various equipment ( coal fueled) | −1.35*** (0.04) | 0.23*** (0.01) | 0.93 (0.09) | 675.12 (0.001) |

*Note*: ***Coefficient β is significant at 1‰; Explanatory variable is *log* Average scale of steam-powered plants (Host technology *H*)

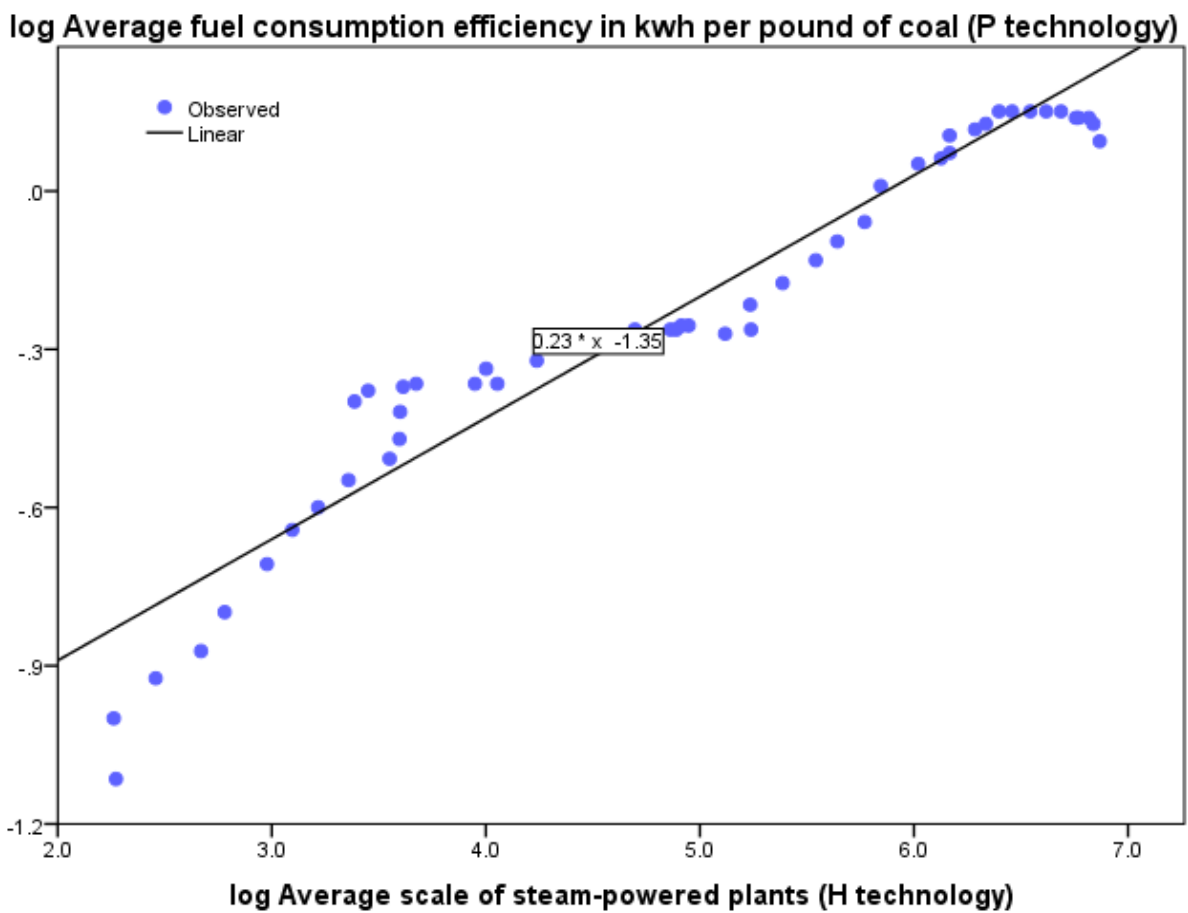

**Figure 3.** Trend and estimated relationship of the evolution of steam-powered electricity with plants that are fossil (coal) fueled (1920-1970 period in US market)

Table 6 shows results of electricity generation with internal-combustion plants having gas turbines; the coefficient of evolutionary growth of this technology is B = 0.35, i.e., B < 1. In short, the evolu-tion of technology in the generation of electricity both in steam-powered plants and internal-combustion plants is low and driven by an evolutionary route of underdevelopment over the course of time (see, Figure 3 and 4).

Table 6 – Estimated relationship for internal-combustion plants with gas turbines (1920-1970 period in US market)

| *Dependent variable*: *log* Average fuel consumption efficiency in kwh per cubic feet of gas (*P*=technological advances of turbine and various equipment) | | | |
|---|---|---|---|
| | Constant $\alpha$ *(St. Err.)* | Evolutionary coefficient $\beta$=B *(St. Err.)* | $R^2$ adj. *(St. Err. of the Estimate)* | F *(sign.)* |
| Gas turbine and various equipment | –2.93*** | 0.35*** | 0.81 | 213.63 |
| | (0.02) | (0.02) | (0.14) | (0.001) |

*Note*: ***Coefficient β is significant at 1‰; Explanatory variable is *log* Average scale of internal-combustion plants (Host technology *H*)

In general, the evolution of technology in the generation of electricity is associated with available natural resources (fossil and gas), the increase in steam pressure and temperature made possible by advances in metallurgy, the use of double reheat units and improvements in the integrated system man-machine interactions to optimize the operation of overall plants, etc. (cf., Sahal, 1981, pp. 183ff). Low rate of technological evolution in the electricity generation technology (underdevelopment with B<1 in tables 5-6) can be due to: "the deterioration in the quality of fuel and of constraints imposed by environmental conditions….other main reasons: First, increased steam temperature requires the use of more costly alloys, which in turn entail maintenance problems of their own…. Thus there has been a decrease in the maximum throttle temperature from 1200 °F in 1962, to about 1000 °F in 1970. Second, there has been lack of motivation to increase the efficiency in the use of gas in both steam-powered and internal-combustion plants because of the artificially low price of fuel due to Federal Power Commission's wellhead gas price regulation. Finally, … there has been a slowdown in generation efficiency due to heavy use of low-efficiency gas turbines necessitated by delays in the construction of nuclear power plant" (Sahal, 1981, p. 184).

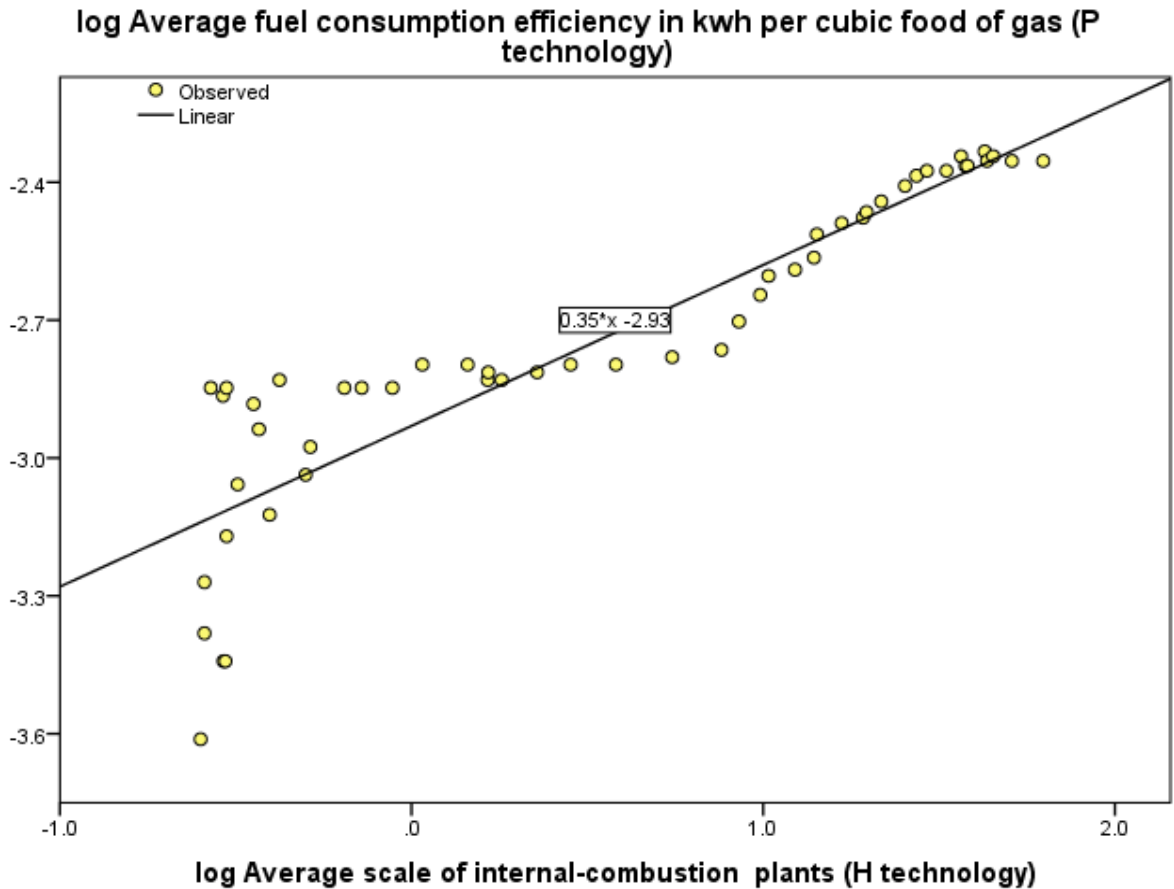

**Figure 4.** Trend and estimated relationship of the evolution of internal-combustion plants with gas turbines (1920-1970 period in US market)

*Results of the evolution of smartphone technology (2008-2018 period in Italian market)*

Table 7 shows that the evolutionary coefficient of growth of smartphone technology is $B = 1.19$, i.e., $B >1$. Technical characteristics of main camera (Parasitic technology $P$) have a disproportionate technological growth in comparison with overall smartphone (Host technology $H$). This coefficient indicates a high grade of the evolution of camera technology supporting a development of complex system of smartphone technology (cf., Figure 5).

Table 7 – Estimated relationship for smartphone technology (2008-2018 period in Italian market)

| *Dependent variable*: | *log* Main Camera in megapixel (*P* technology) | | | |
|---|---|---|---|---|
| | *Constant* $\alpha$ *(St. Err.)* | *Evolutionary coefficient* $\beta=B$ *(St. Err.)* | $R^2$ *adj.* *(St. Err.* *of the Estimate)* | *F* *(sign.)* |
| Main Camera | 2.07*** | 1.19*** | 0.97 | 897.483 |
| technology | (0.03) | (0.04) | (0.18) | (0.001) |

*Note*: ***Coefficient β is significant at 1‰; Explanatory variable is *log* Processor GHz (technological advances of smartphone–Host technology *H*)

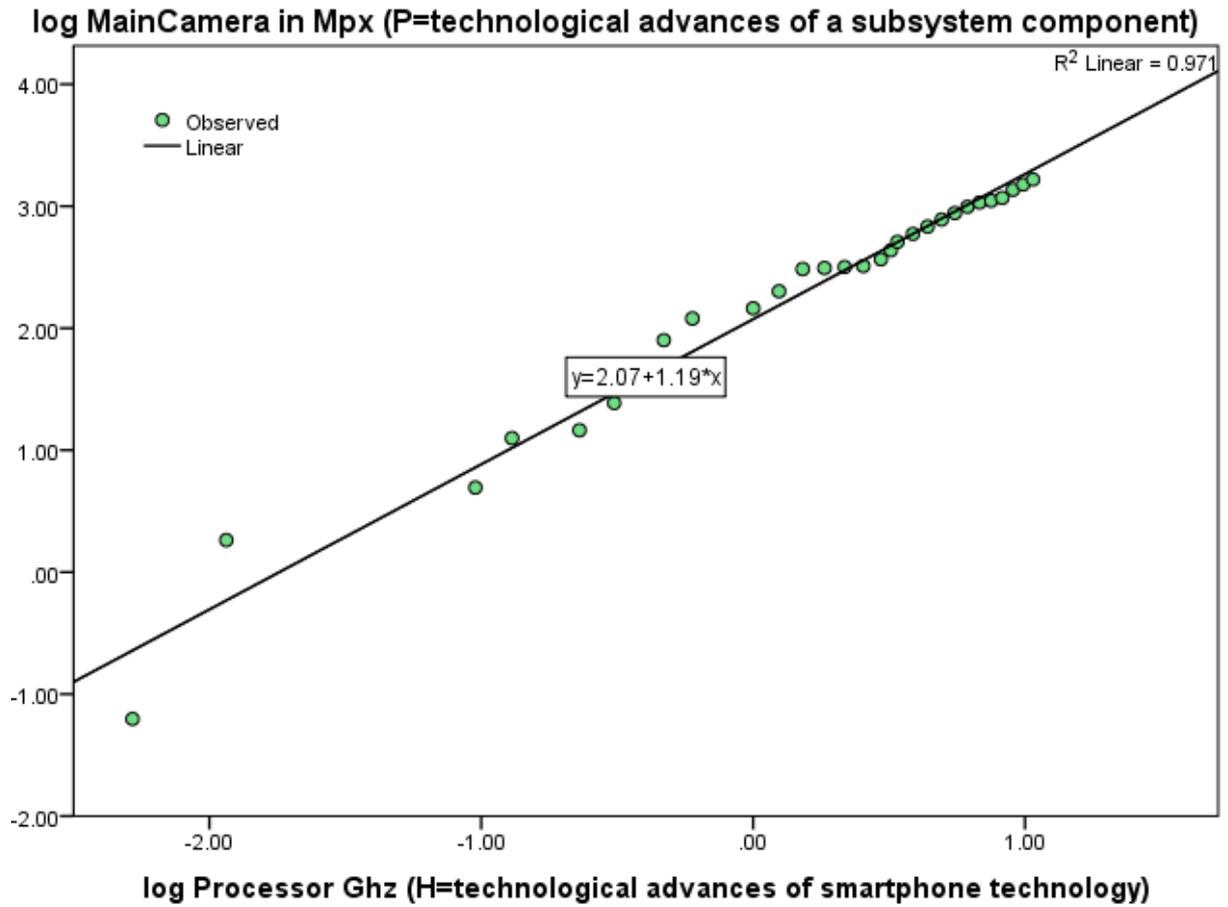

**Figure 5.** Trend and estimated relationship of the evolution of main camera in smartphone technology (2008-2018 period in Italian market)

Table 8 – Estimated relationship for the evolution of smartphone technology considering multidimensional interaction between host system and subsystems of parasitic technologies (*log-log* model, 2008-2018 period in Italian market)

*Dependent variable*: *log* Main Camera in megapixel ($P_1$ technology) at *t* =2008, …, 2018

| Smartphone | Unstandardized Coefficient | Standardized Coefficient | t-test |
|---|---|---|---|
| Constant. α | −1.19 | | −1.83 |
| (St. Err.) | (0.65) | | |
| | | | |
| Predictors ⇓ | | | |
| Coefficient *log* $P_2$ technology 2nd Camera  megapixel | 0.09*** | 0.17 | 4.65 |
| (St. Err.) | (0.02) | | |
| | | | |
| Coefficient  *log* $P_3$ technology Resolution Display in pixels | 0.14*** | 0.19 | 4.12 |
| (St. Err.) | (0.03) | | |
| | | | |
| Coefficient  *log* $P_4$ technology RAM Gb | 0.20*** | 0.24 | 3.84 |
| (St. Err.) | (0.05) | | |
| | | | |
| Coefficient  *log* $P_5$ technology Memory Gb | 0.12*** | 0.20 | 4.38 |
| (St. Err.) | (0.03) | | |
| | | | |
| Coefficient  *log* $P_6$ technology  Battery mAh | 0.14* | 0.07 | 1.97 |
| (St. Err.) | (0.07) | | |
| | | | |
| Coefficient  *log* H technology  Processor GHz | 0.12 | 0.06 | 1.46 |
| (St. Err.) | (0.08) | | |
| | | | |
| $R^2$ *adj. adj.* | 0.70 | | |
| *(St. Err. of the Estimate)* | (0.29) | | |
| | | | |
| *F* | 233.81 | | |
| *(sign.)* | (0.001) | | |

*Note*: P*i*=Parasitic technology *i*=1, …, 6; H=Host technology (smartphone)
*** *p-value*< .001
** *p-value*< .010
* *p-value*< .050

Table 8 shows that the evolutionary pathways of camera technology in smartphone is mainly driven by advances of RAM in Gb, memory in Gb and display resolution in pixels, as showed by standardized coefficients of regression (see, highlighted cell  in the third column of table 8). $R^2$ adjusted of the model [7] indicates that about 70% of the variation in megapixels of main camera can be attributed (linearly) to predictors indicated in table 8. Figure 6 shows that the coevolution of technical characteristics of host system and parasitic technologies in smartphone technology. Table 9 reveals that main camera has a very high coefficient of correlation with other parasitic technologies and with processor GHz (a proxy of the technical advances of overall smartphone-host technology): in general, *r*>.78 (*p*-value 0.001), except for battery mAh. This result suggests that the evolution of smartphone technology is due to coevolutionary processes of different parasitic technologies in a complex system of technology.

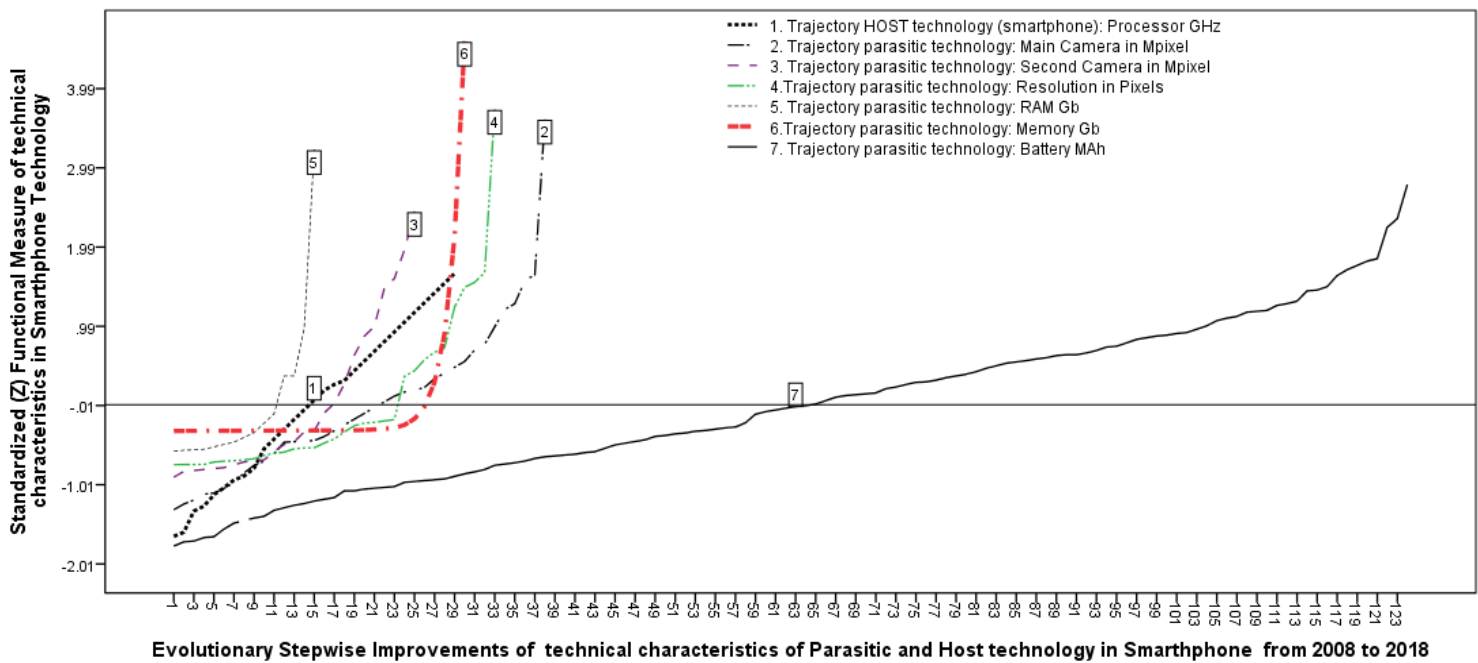

**Figure 6.** Coevolution of technical characteristics of host and subsystem parasitic technologies in smartphone (2008-2018 period).

**Note:** The Functional Measures of Technology *i* in *t* (FMT*i, t*) of *y*-axis are systematized in a comparable framework by applying the following standardization formula for the technology *i* in *t=time*: $Z(FMT)_{it} = \frac{FMT_{it} - \mu_t}{\sigma_t}$; where: $Z(FMT)_{it}$ = standardized FMT$_{it}$ (Functional Measures of Technology *i* at *t*); $FMT_{it}$ = Functional Measures of Technology *i* at the year *t*; $\mu_t$ = arithmetic mean of the FMT over *t*; $\sigma_t$ = standard deviation of the FMT over *t*. *Remark*: $FMT_{it}$ is negative when the raw score is below the arithmetic mean, positive when it is above. A zero value of $FMT_{it}$ indicates that the raw value is equal to the arithmetic mean.

Table 9. Correlation between advances of technical characteristics of main camera, host and other parasitic technologies in smartphone (2008-2018 period)

|  |  | **HOST** *Log* Processor GHz | *Parasitic 2 Log* Second Camera MP | *Parasitic 3 Log* Resolution Pixels | *Parasitic* 4 *Log* RAM Gb | *Parasitic 5 Log* Memory Gb | *Parasitic 6 Log* Battery MAh |
|---|---|---|---|---|---|---|---|
| Log *Parasitic* 1 Main Camera Mpx | Pearson Correlation | .985** | .903** | .929** | .933** | .781** | .295 |
|  | Sig. (2-tailed) | .001 | .001 | .001 | .001 | .001 | .072 |
|  | *N* | 29 | 25 | 33 | 15 | 30 | 38 |

**. Correlation is significant at the 0.01 level (2-tailed). *N*=technical improvements from 2008 to 2018

In particular, the rapid evolution of smartphone technology ($B>1$ in table 7) is due to numerous innovations over time, such as Bluetooth for wireless communication in 2002, touchscreen in 2007, app store and android market in 2008 that have generated many parasitic technologies given by software applications for mobile devices, Siri and fingerprint scanners in 2011, 4G in 2012, waterproof phone in 2013, dual camera in 2014, 4K HDR resolution display in 2015, modular phones in 2016, and facial recognition in 2017, etc. This finding indicates that the long-run evolution of smartphone technology depends on the behavior and coevolution of inter-related parasitic technologies (cf., Coccia, 2018). Moreover, learning effects, based on learning by doing and learning by using, are fostering the assimilation of new technology in smartphone devices from many parasitic technologies to support the evolutionary pathway of overall complex system of technology (Cohen and Levinthal, 1990). Sahal (1981, p. 82, original italics) argues that: "the role of learning in the *evolution* of a technique has profound implications for its *diffusion* as well". In the context of smartphone technology, Watanabe et al. (2012, pp. 1293-1294) argue that learning effects in ICTs can be the sources of its self-propagating development of technology, acquiring new functionality from digital industry, wireless communications and software applications (cf., Carranza, 2010; Coccia, 2018).

Overall, then, this statistical analysis shows that the proposed models here can assist in assessing explaining the evolution of different technologies based on interaction between host system and its subsystem of technology that guides evolutionary pathways and technological diversification over time and space (cf., Coccia, 2018).

**Discussion and conclusion**

Many characteristics in the nature and evolution of technology are hardly known. Scientists should open the debate regarding the nature and types of interaction between host technologies and its subsystem technologies that may explain and generalize aspects of the evolution of technology and technical change in society (cf., Coccia, 2018; Pistorius and Utterback, 1997; Sandén and Hillman, 2011). Some scholars argue that technologies and technological change display numerous life-like features, suggesting a deep connection with biological evolution[7]. The analogy between biological processes and technological evolution is a source of ideas because biological evolution has been studied in-depth and provides a logical structure of scientific inquiry for the evolution of technology.

This study applies a broad analogy between evolutionary ecology of parasites and technological evolution, within a theoretical framework of Generalized Darwinism, to propose a theory to measure, assess and predict the evolutionary pathways of technology. The evolution of technology here is based on an assumption that technologies are complex systems that interact in a nonsimple way with other technologies and inter-related subsystems of technology. In particular, this study analyses the evolution of technology considering the interaction between host technology (system) and parasitic technology (subsystem). The approach here is operationalized with a simple model that contains only two parameters and provides the coefficient of evolutionary growth, which is useful to measure and assess the effect that host technology can have on parasitic technology's growth rate, predicting which technologies are likely to evolve rapidly. The technometrics here suggests three simple grades of the evolution of technology, based on the coefficient of evolutionary growth, according to host technology $H$ can enhance or inhibit the growth rate of parasitic technology $P$: B<1 (underdevelopment of $P$), $B$=1 (growth of $P$) and $B$>1 (development of $P$ and of the whole system of technology). The proposed technometrics, tested in five example technologies, provides consistent results of the evolution of technologies with empirical data and the history of specific technologies under study.

In general, the evolution of technology has *universals* based on mutualistic and symbiotic interaction, similar to many phenomena in nature and society. In fact, Szathmáry (2011) argues that benefits of cooperation can drive the evolution of a system that supports cooperative behavior. Technological interaction based on cooperation between technologies (e.g., mutualism and symbiosis) must pay off in the long run, even if it is immediately costly to cooperative technologies due to switching costs for adapting to evolving host technology (e.g., the transition of headphones from wired to wireless technology with new generations of electronic devices without jack) .

Coefficient of evolutionary growth B here is a metric for classifying the modes of technological interaction and for predicting the long-term development of complex system of technology, namely:

1. Coefficient B<1 suggests low interaction between host system and its subsystem of technology (technological parasitism), whereas B>1 suggests a high interaction between host system and subsystem of technology (technological symbiosis).

2. Technology having an accelerated growth of its parasitic technologies (B>1) advances rapidly, whereas technology with low growth of its parasitic technologies (B<1) enhances slowly.

3. High development of technology is governed by a process of disproportionate growth in its parasitic technologies (B>1), such as the technological development of farm tractor, smartphone and freight locomotive technologies described here.

4. Evolution of technology is inhibited when its parasitic subsystem P has low changes in relation to changes of host technology (B<1), generating underdevelopment of the whole system of technology over the course of time (e.g., the generation of electricity in steam-powered and internal-combustion plants).

5. Long-run evolution of a technology depends on the behavior and evolution of associated parasitic technologies. To put it differently, long-run evolution of a specific technology is enhanced by the integration of two or more parasitic/symbiotic technologies that generate co-evolution of the overall complex system of technology.

Overall, then, one of the most important findings of the proposed theoretical framework here is two general properties of the evolution of technology as a complex system:

(a) the disproportionate growth of technological subsystems in a host technology generates the development of overall complex system of technology

(b) Interaction between technologies can generate coevolution within complex system of technology with the shift from technological parasitism (indicated with B<1) to technological symbiosis (B>1) over the course of time. This transition dynamics is due to *natural selection* of technical characteristics during the interaction between technologies that reduces negative effects and favors positive effects directed to an evolution of reciprocal adaptations of technologies in complex systems of technology over time and space (cf., property of mutual benefaction by Coccia, 2018).

The finding of this study could aid policymakers and managers to design best practices of technology policy and management of technology for supporting development of new technology, and as a consequence, industrial and economic change in society. One of the main limitations of this approach is the lack of useful data in sufficient quality for different technologies. Future efforts in this research field require a gathering of substantial amount of technological characteristics for different technologies to provide further empirical evidence of the evolutionary pathways of technology over time and space. Moreover, future study will be also directed to support the theory here with practical policy and management implications to guide funding for R&D towards specific technologies (having B>1) that are likely to evolve rapidly in society.

Overall, then, the idea presented in the study here to measure, analyze and predict evolution of technology is adequate in some cases but less in others because of the diversity of technological innovations and their relationships in different complex systems and socioeconomic environments. Nevertheless, the broad analogy between evolutionary ecology of parasites and technological evolution, based on Generalized Darwinism, keeps its validity here in explaining and predicting general evolutionary pathways of technology. In particular, the proposed approach here based on the ecology-like interaction between technologies—may lay the foundation for development of more sophis-ticated concepts and theoretical frameworks in technometrics and technological forecasting. As a matter of fact, these findings here can encourage further theoretical and empirical exploration in the *terra incognita* of the interaction between different technologies during economic change to measure, explain and predict the aspects of the evolution of technology. To conclude, this study constitutes an initial significant step in measuring the evolution of technology considering the interaction between technologies in complex systems to predict the long-run behavior of technology in society. However, the identification of a comprehensive technometrics for technological forecasting in different domains of technology, having a technological diversification in markets, is a non-trivial exercise. In fact, Wright (1997, p. 1562) properly claims that: "In the world of technological change, bounded rationality is the rule."

Acknowledgement

I wish to thank Sara I. Walker, Josh Watts, Paul Davies¸ Michael Barton (Arizona State University), Trang T. Thai (Intel Corporation), seminar participants at the Beyond-Center for Fundamental Concepts in Science (ASU in Tempe, AZ), and especially four anonymous referees for valuable comments and discussions. I am grateful to Diego Margon for accurate research assistance. I also thank the Library of Arizona State University for scientific material provided on these topics. All remaining errors are my own. Financial support by National Endowment for the Humanities and National Research Council of Italy–Direzione Generale Relazioni Internazionali (Research Grant n. 0072373-2014 and n. 0003005-2016) for my visiting at Arizona State University where this research started in 2014 is gratefully acknowledged. Usual disclaimer applies.

**Appendix**

*Table 1A* – Descriptive statistics in *log* scale

| | *log*<br>Fuel consumption efficiency in horsepower hours<br>(Engine of Tractor P) | *log*<br>Mechanical efficiency ratio of draw-bar horsepower to belt<br>(Tractor efficiency H) | *log*<br>Tractive efforts in pound<br>(Locomotive power P) | *log*<br>Total railroad mileage<br>(Infrastructure for locomotive H) |
|---|---|---|---|---|
| Years | 44 | 44 | 29 | 29 |
| Mean | 2.13 | 4.19 | 10.43 | 12.86 |
| Std. Deviation | 0.27 | 0.146 | 0.22 | 0.11 |
| Skewness | -0.76 | -0.68 | -0.21 | -1.04 |
| Kurtosis | -0.83 | -0.56 | -1.19 | -0.06 |

| | *log*<br>Average fuel consumption efficiency in kwh per pound of coal<br>(turbine and various equipment in steam-powered plants P) | *log*<br>Average scale of steam-powered Plants<br>H | *log*<br>Average fuel consumption efficiency in kwh per cubic feet of gas<br>(turbine and various equipment in internal-combustion plants P) | *log*<br>Average scale of internal-combustion plants<br>H |
|---|---|---|---|---|
| Years | 51 | 51 | 51 | 51 |
| Mean | -0.25 | 4.85 | -2.75 | 0.51 |
| Std. Deviation | 0.34 | 1.43 | 0.33 | 0.85 |
| Skewness | -0.67 | -0.17 | -0.67 | 0.02 |
| Kurtosis | -0.09 | -1.26 | 0.04 | -1.64 |

| | *log*<br>Main Camera mega-pixel in smartphone $P_1$ | *log*<br>Processor Giga Hertz in smartphone H | *log*<br>Second Camera mega-pixel in smartphone $P_2$ | *log*<br>Memory Giga byte in smartphone $P_3$ |
|---|---|---|---|---|
| Years | 10 | 10 | 10 | 10 |
| Mean | 2.54 | 0.13 | 1.43 | -0.31 |
| Std. Deviation | 2.80 | 0.41 | 1.39 | -1.09 |
| Skewness | -1.52 | -1.38 | -0.13 | 0.84 |
| Kurtosis | 3.05 | 1.65 | -0.88 | 0.51 |

| | *log*<br>RAM Giga byte in smartphone $P_4$ | *log*<br>Battery milliAmpere hour in smartphone $P_5$ | *log*<br>Display resolution total pixels in smartphone $P_6$ |
|---|---|---|---|
| Years | 10 | 10 | 10 |
| Mean | 0.30 | 7.64 | 13.12 |
| Std. Deviation | 0.41 | 7.77 | 13.33 |
| Skewness | -0.16 | -6.94 | -0.50 |
| Kurtosis | -0.65 | 64.64 | -0.55 |

*Note*: P=parasitic technology; H= Host technology. Numbers *x* in table are in natural logarithmic and have to be transformed with $e^x$ to obtain absolute value.

Biographical note

Mario Coccia is a social scientist at the National Research Council of Italy (CNR) and visiting scholar at YALE university. He has been research fellow at the Max Planck Institute of Economics and visiting professor at the Polytechnics of Torino and University of Piemonte Orientale (Italy). He has conducted research work at the Arizona State University, Georgia Institute of Technology, United Nations University-Maastricht Economic and Social Research Institute on Innovation and Technology (UNU-MERIT), RAND Corporation (Washington D.C.), University of Maryland (College Park), Bureau d'Économie Théorique et Appliquée (Strasbourg), Munk School of Global Affairs (University of Toronto), and Institute for Science and Technology Studies (University of Bielefeld). He leads CocciaLAB at CNR to investigate, with interdisciplinary methods of inquiry, the determinants of socioeconomic phenomena, such as terrorism, war, crime, new technology, evolution of scientific fields and technologies, human progress, etc. He has written more than two hundred-fifty papers in several disciplines.